\documentclass[11pt,english]{article}
  \usepackage[T1]{fontenc}
  \usepackage{lmodern}
  \usepackage[latin9]{inputenc}
  \usepackage{geometry}
  \usepackage{amsthm}
  \usepackage{amsmath}
  \usepackage{amssymb}
  \usepackage{array}
  \usepackage{enumitem}
  \usepackage{graphicx}
\usepackage[activate={true,nocompatibility},final,tracking=true,kerning=true,spacing=true,factor=1100,stretch=10,shrink=10]{microtype}
  \usepackage{setspace}
  \usepackage{tabu}
  \usepackage[font={small}]{caption}
  \usepackage{color}
  \usepackage{braket}
  \usepackage{microtype}
  \usepackage{verbatim}
  \usepackage{fancyhdr}
  \usepackage{titling}
    \usepackage[affil-it]{authblk}
  \usepackage[round]{natbib}
  \def\dbar{{\mathchar'26\mkern-12mu d}}
\begin{document}
 
  \title{Black Hole Entropy is Thermodynamic Entropy}
	\author[1,2]{Carina E. A. Prunkl\thanks{\texttt{carina.prunkl@philosophy.ox.ac.uk}}
  } 
  	\author[3]{Christopher G. Timpson\thanks{\texttt{christopher.timpson@bnc.ox.ac.uk}} 
}   
 \affil[1]{Future of Humanity Institute, University of Oxford}
 \affil[2]{Black Hole Initiative, Harvard University}
\affil[3]{Brasenose College, University of Oxford}

  \maketitle

 \begin{abstract}
 The comparison of geometrical properties of black holes with classical thermodynamic variables reveals surprising parallels between the laws of black hole mechanics and the laws of thermodynamics. Since Hawking's discovery that black holes when coupled to quantum matter fields emit radiation at a temperature proportional to their surface gravity, the idea that black holes are genuine thermodynamic objects with a well-defined thermodynamic entropy has become more and more popular. Surprisingly, arguments that justify this assumption are both sparse and rarely convincing. Most of them rely on an information-theoretic interpretation of entropy, which in itself is a highly debated topic in the philosophy of physics. We discuss some of the pertinent arguments that aim at establishing the identity of black hole surface area (times a constant) and thermodynamic entropy and show why these arguments are not satisfactory. We then present a simple model of a Black Hole Carnot cycle to establish that black hole entropy is genuine thermodynamic entropy which does not require an information-theoretic interpretation.
 \end{abstract}
 \newpage
 
 \tableofcontents
  \newpage
  
 \section{Introduction}
  
\begin{quotation}
\small   I believe that in order to gain a better understanding of the degrees of freedom responsible for black hole entropy, it will be necessary to achieve a deeper understanding of the notion of entropy itself. Even in flat spacetime, there is far from universal agreement as to the meaning of entropy --- particularly in quantum theory --- and as to the nature of the second law of thermodynamics. The situation in general relativity is considerably murkier [...]. \citep[p.27]{wald_thermodynamics_2001}
\end{quotation}

In the last few decades, black holes have enjoyed an increasing amount of attention from physicists and philosophers alike, not least because of their exceptional status as objects for whose full description general relativistic, quantum theoretic and thermodynamic considerations seem to be needed. The surprising parallels between geometrical properties of black holes and classical thermodynamic variables have been taken to suggest that there exists a deeper connection between the laws of black hole mechanics and the laws of thermodynamics. A connection that might go beyond mere analogy, possibly allowing us to \textit{identify} the respective geometrical properties with their thermodynamic counterparts. This would then imply that black hole entropy, proportional to the black hole's event horizon area, is in fact identical to the thermodynamic entropy. With this identification remaining unchallenged by most physicists, a remarkable amount of effort is instead spent on finding a microphysical underpinning of the Bekenstein-Hawking (black hole) entropy, and with it an appropriate interpretation. However, until now no agreement on the matter is in sight \citep{wald_black_1993,susskind_speculations_1993,bombelli_quantum_1986,frolov_black_1998,bekenstein_bekenstein-hawking_2008} and so the pressing question remains: what is black hole entropy?

To make things worse, there is a remarkable amount of disagreement about the meaning of entropy even in cases where we \textit{do} have a firm grip on the microphysics. Even if we deal with standard thermodynamic scenarios that do not involve black holes, it is far from clear which, if any, statistical entropy ought to be taken as the correct statistical mechanical generalisation of the phenomenological entropy, as was shown in the previous chapter. In addition to being conceptually distinct, the statistical mechanical candidates may even numerically disagree. In particular, none of them manages to recover all of the properties that are usually deemed necessary for the phenomenological entropy, as shown in  \citep{callender_reducing_1999} for the case of Gibbs and Boltzmann entropies. And so, even if one were to find an unproblematic microphysical underpinning of the Bekenstein-Hawking expression, it would be far from clear whether this in fact would resemble thermodynamic entropy.

In the face of the amount of disagreement about the nature of entropy on the one hand, and the growing importance of black hole thermodynamics for the foundations of physics on the other hand, we want to contribute to the debate about the nature of black hole entropy by showing that, within a given range of external parameters, black holes can in fact be considered genuine thermodynamic objects.\footnote{A recent analysis by \cite{wallace_case_2017} examines the same question from a slightly different angle, but comes to the same conclusion.} We will not do so by presenting yet another statistical mechanical argument for why entropy is proportional to horizon area, for the aforementioned reasons. Instead, we are interested in the question of whether we can consider the Bekenstein-Hawking entropy to be a \textit{thermodynamic} entropy in the first place. We will therefore try to use only phenomenological reasoning, thereby avoiding statistical arguments as far as possible. ``As far as possible'' because we take as given Hawking's \citeyear{hawking_particle_1975} result that black holes emit radiation at a temperature proportional to their surface gravity. This derivation of the Hawking radiation can of course not be done phenomenologically, but requires quantum field theory applied to curved space-time. It has hence ultimately a statistical (or rather quantum) origin. However, once established, the Hawking radiation behaves just like ordinary thermal radiation\footnote{We are only concerned with an electromagnetic field in the vicinity of the black hole, but suspect that our reasoning can be extended to imply other quantum fields as well.}, whose behaviour uncontroversially can be described in a phenomenological thermodynamic setting. 

The argument for a black hole entropy presented here furthermore differs from previous attempts, as it avoids controversial concepts such as `information' and does not rely on the identification of a statistical mechanical entropy with the thermodynamic entropy. Instead, the traditional notion of a Carnot cycle is be used to derive the expression for the Bekenstein-Hawking entropy. This Carnot cycle is implemented by considering a box containing a black hole and a photon gas that undergo a series of quasi-static thermodynamic transformations. By doing so, it will be shown that when a black hole is coupled to an uncontroversially thermodynamic object, i.e. the photon gas, the two of them combined behave thermodynamically and moreover entropy differences are proportional to differences of horizon area. This will provide strong reason to believe that black hole entropy is indeed thermodynamic entropy.

The idea of putting black holes into boxes filled with radiation has been previously proposed, most prominently by \cite{hawking_black_1976} and later by \cite{custodio_thermodynamics_2003}. Both established that stable equilibrium states between a black hole and a photon gas exist within a certain range of external parameters. However, in both approaches the argument is made by presupposing and appealing to the (statistical) entropy, which is avoided in the present discussion. A Carnot cycle cycle involving a black hole in a box has furthermore been suggested by \cite{opatrny_black_2011}, where the authors use two black holes as heat sources/heat sinks respectively, therefore varying from this approach where the black hole is taken to be part of the working medium. 

We will begin with a short recap of the laws of black hole mechanics before revisiting some of the arguments that have previously been made in an attempt to establish that black holes have an entropy that is proportional to their respective horizon areas. The two most prominent accounts are given by \cite{bekenstein_black_1973} and \cite{hawking_black_1976}. In the second part of the article, we will then present the black hole Carnot cycle and derive the expression for the thermodynamic entropy.

\section{Preliminaries - Laws of Black Hole Mechanics}

In this section we will briefly discuss the similarities and differences between the laws of thermodynamics and the laws of black hole mechanics. A summary of the laws can be found in Table \ref{table}. 

\begin{table}
\begin{center}
\def\arraystretch{1.5}
\begin{tabular}{| c | m{.35\textwidth} | m{.35\textwidth} |}
	\hline
	  & \textbf{Thermodynamics} & \textbf{Black Hole Mechanics} \\
     \hline
 	\textsc{Zeroth Law} & Two systems in thermal equilibrium with a third are in thermal equilibrium with each other. & $\kappa$ is constant across the event horizon.  \\  
    \hline
 	\textsc{First Law} & \begin{equation}dU=\dbar Q- \dbar W\nonumber\end{equation} & \begin{equation}dM=\frac{1}{8\pi}\kappa dA+\Omega dJ \nonumber\end{equation} \\
	\hline
    \textsc{Second Law} & \begin{equation}dS_{TD}\geq 0\nonumber\end{equation} & \begin{equation}d\left(S_{exterior}+\frac{c^3}{4G\hbar}A\right)\geq 0\nonumber\end{equation} \\
    \hline
    \textsc{Third Law} & \begin{equation} dS_{TD}\rightarrow 0 \;\text{as}\; T\rightarrow 0\nonumber \end{equation} & It is impossible to achieve $\kappa = 0$ within a finite number of steps. \\
    \hline
\end{tabular}
\caption{Summary of the laws of thermodynamics and their black hole mechanical counterparts.}
\label{table}
\end{center}
\end{table}



The \textbf{zeroth law} of black hole mechanics states that the surface gravity $\kappa$ of a static black hole is constant over the whole event horizon \citep{bardeen_four_1973}. The surface gravity of a black hole is given by the proper acceleration of a test particle near the event horizon multiplied by a redshift factor. The fact that the surface gravity is constant is said to resemble the zeroth law of thermodynamics, often expressed as the statement that \textit{temperature} is constant throughout a body in thermal equilibrium. Traditionally, however, the zeroth law is concerned with the \textit{transitive} relation between equilibrium states. It is merely a consequence of this transitive relation that the temperature of a system in thermal equilibrium is constant throughout the system itself. The zeroth law of black hole mechanics as described above therefore merely recovers a consequence of its thermodynamic counterpart, cheekily ignoring the far more challenging task of establishing transitive equilibrium.

The \textbf{first law} of black hole mechanics states that any change in black hole mass $M$ needs to be balanced by a change in either surface area $A$ and/or angular momentum $J$:
\begin{equation}
dM=\frac{1}{8\pi}\kappa dA+\Omega dJ,\label{eq:firstlawBH}
\end{equation}

where $\Omega$ is called the \textit{angular velocity of the horizon} \citep{wald_thermodynamics_2001} and is constant. An extra term including an electrostatic potential and a change in charge may be added to the equation, but is of no importance here. All of the above entities are defined for an observer at infinity, unless stated otherwise. 

Equation (\ref{eq:firstlawBH}) is taken to resemble the first law of thermodynamics,

\begin{equation}
dU= TdS-\dbar W, \label{eq:firstlawTD}
\end{equation}

where $dU$ resembles the change in energy, $T$ the temperature, $dS$ the change in entropy, and $\dbar W$ the work done by the system. Normally one encounters $\dbar Q$, and not $TdS$ for the heat term. This is, because strictly speaking, $T$ and $S$ do not have their physical meaning as absolute temperature and entropy at this point, but are merely mathematical placeholders used to convert the inexact differential into an exact one. It is only by the second law that they obtain their canonical meaning as absolute temperature and entropy.

The resemblance between equation (\ref{eq:firstlawBH}) and equation (\ref{eq:firstlawTD}) (the surface area sits in Equation (\ref{eq:firstlawBH}) just as entropy sits in (\ref{eq:firstlawTD})), plus the fact that black hole horizon area originally was taken to be non-decreasing, led \citeauthor{bekenstein_black_1973} in 1973 to express his suspicion that the black hole horizon area could be interpreted as playing the role of the thermodynamic entropy\footnote{It is widely accepted that for a non-evaporating black hole a straightforward identification of the laws of black hole mechanics with the laws of thermodynamics fails as the temperature of such a black hole is necessarily zero at all times. See however \cite{curiel_classical_2014} for an opposing view on the matter. In his article, Curiel claims that even classical black holes may be considered to be thermodynamic objects with a temperature.}. He proposed the \textbf{generalised second law} of black hole mechanics: the change of entropy in the black hole exterior plus the change of black hole surface area must be non-negative. At the time of Bekenstein's proposal there still existed a problem about identifying the non-zero surface gravity of a black hole with what was thought to be a zero thermodynamic temperature. This issue was resolved when \citet{hawking_particle_1975} discovered that black holes are in fact not black but instead emit radiation at a temperature proportional to their surface gravity. It was Hawking's result which led to a precise definition of the black hole entropy as $S_{BH}=\frac{c^3A}{4G\hbar}$.  The generalised second law of black hole mechanics \citep{bekenstein_black_1973,hawking_black_1976} then reads
\begin{equation}
\delta\left(S_{exterior}+\frac{c^3}{4G\hbar}A\right)\geq 0,
\end{equation}

where $S_{exterior}$ refers to the entropy of the black hole exterior\footnote{Curiously, \cite{bekenstein_black_1973} calls this the `common entropy' in the black hole exterior. It will be shown later, that `common', for Bekenstein, does not mean `thermodynamic'.}. 

The third law of thermodynamics states that the entropy change of a system undergoing a reversible isothermal process approaches zero as the temperature at which that process is performed approaches zero kelvin. A consequence of this is that it is impossible to reduce the entropy of an object to zero within a finite number of steps. The \textbf{third law} of black hole mechanics is then the statement that it is impossible to achieve a surface gravity of zero within a finite number of steps. It will however play no role for the argument presented here.


\section{Which Entropy?}

\subsection{Black Hole Entropy in the Literature}
It is certainly undeniable that the laws of thermodynamics and the laws of black hole mechanics bear some extraordinary resemblance. However, in order to show that the laws of black hole mechanics simply \textit{are} the laws of thermodynamics applied to black holes, more work needs to be done. In particular one must show that the entities referred to are in fact the same in both cases. This means demonstrating that some of the geometrical properties of black holes can indeed be identified with the thermodynamic variables. Here we will only consider the correspondence of the thermodynamic entropy with the black hole horizon area. As pointed out before, we will take it as uncontroversial that Hawking's \citeyear{hawking_particle_1975} result about particle creation at the black hole event horizon is correct, which allows us to identify the temperature of the electromagnetic quantum field outside the horizon with the ordinary black body temperature. 

We will begin with a brief discussion of some of the already existing arguments. Both the original argument made by \cite{bekenstein_black_1973} and Hawking's follow-up in 1976 make use of the concept of `information' in order to demonstrate that black holes have an entropy. It is debatable whether their arguments aim indeed at demonstrating that black hole entropy is thermodynamic---as opposed to merely a statistical mechanical---entropy. We will come back to this issue shortly, but for now we will just follow the widespread opinion that this is the case. W\"uthrich \citeyear{wuthrich_are_2017} and \cite{dougherty_black_2016} for example take the establishing of $S_{BH}=S_{TD}$ to be a central part of Bekenstein's and, in the latter case, also Hawking's work and indeed, many physicists take black hole entropy to be of a thermodynamic nature \citep{page_information_1993,wald_thermodynamics_2001,braunstein_better_2013}.

Before giving a more detailed account of the existing arguments, however, it is useful to distinguish between three routes one might chose to show that black holes are genuine thermodynamic objects with a thermodynamic entropy proportional to their surface area. They are

\begin{enumerate}[align=left]
\item[\textsc{Similarity:}] The laws of black hole mechanics have the same structure as the laws of thermodynamics. In particular, surface area is related to mass in the same way entropy is related to internal energy. 
\item[\textsc{Preservation:}] An omission to identify thermodynamic entropy with black hole horizon area leads to a violation of the second law of thermodynamics. 
\item[\textsc{Statistical Mechanics:}] It is possible to derive a statistical mechanical entropy that is proportional to black hole surface area. 
\end{enumerate}

As we can see, \textsc{Similarity} appeals to the similarities in behaviour of horizon area and thermodynamic entropy, \textsc{Preservation} to the preservation of the second law and \textsc{Statistical Mechanics} to the possibility of deriving an entropy from statistical grounds. The above strategies are related, of course, and neither of them would by itself be a convincing reason to believe in a thermodynamic black hole entropy. \textsc{Similarity}, for example, can at most establish an analogy---but not an identity---between black hole mechanics and thermodynamics, whereas \textsc{Preservation} actually requires this identity. Nevertheless, \textsc{Similarity} lays the conceptual groundwork for why one should suspect that there is any connection between entropy and horizon area in the first place. \textsc{Preservation} then provides the actual physical link between black hole mechanics and thermodynamics. \textsc{Statistical Mechanics} delivers the mathematical underpinning for it, but requires a prior commitment to which (if any) statistical mechanical entropy the thermodynamic entropy is reducible to, a highly controversial matter \citep{callender_reducing_1999,sklar_reduction?_1999,goldstein_boltzmanns_2001,wallace_non-problem_2013}. 

The above distinctions will make it easier to identify the lines of argument chosen by Bekenstein and Hawking to establish black hole entropy, as will be now shown.

\subsubsection*{Bekenstein}
We will begin with with a recapitulation of the argument given by Bekenstein, who in his influential \citeyear{bekenstein_black_1973} article attempted a ``unification of black hole physics and thermodynamics'' \citep[p.2334]{bekenstein_black_1973}. He begins his argument by pointing out some similarities between black hole area and thermodynamic entropy: the energy change of a black hole is as intimately related to a change of horizon area, as is (internal) energy change to a change of entropy in the thermodynamic case. Furthermore, a merging of Schwarzschild black holes allows for the extraction of energy in the form of gravitational waves. 
Analogously, two thermodynamic systems, each individually at equilibrium, allow for work extraction when brought into thermal contact. These observations all belong to \textsc{Similarity}.

So do the following ones, although here the similarities are taken to be in the relationship between entropy and information and so are based on a particular statistical mechanical generalisation of the thermodynamic entropy (and a particular interpretation of the former). Bekenstein takes entropy to be best understood in terms of inaccessible information about the system's degrees of freedom:

\begin{quote}\small\singlespacing
The connection between entropy and information is well known. The entropy of a system measures one's uncertainty or lack of information about the actual internal configuration of the system. \citep[p.2335] {bekenstein_black_1973}
\end{quote}

Note, however, that for Bekenstein, information-theoretic entropy is a more general notion than thermodynamic entropy, even though the latter is to be understood in information-theoretic terms. A system's thermodynamic entropy, according to Bekenstein, measures the amount of uncertainty about the system's internal configuration, given few macroscopic parameters such as temperature, volume or pressure. In the case of black holes, this uncertainty allegedly is of a deeper kind. Similarly to the thermodynamic case, however, it is possible to characterise the state of the black hole by a small number of parameters: the black hole mass, its charge and angular momentum \citep{misner_gravitation_1973}. Furthermore, given that there are numerous ways the black hole could have formed\footnote{Note the difference between the normal case, in which the uncertainty refers to actual possible realisations, and the black hole case, in which the uncertainty refers to \textit{past} histories of how the black hole has formed.}, there must exist an inaccessible multiplicity of states for each set of parameters. This resembles the thermodynamic case, where each combination of macroscopic parameters can be realised by multiple configurations on the microscopic level. Entropy is then said to quantify this multiplicity. To give an (information theoretic) intuition of why entropy is related to surface area, Bekenstein considers radiation or particles that fall into a black hole: information about the infalling objects is lost, but at the same time the surface area of the black hole increases, quantifying an increase in ignorance.

The above reasonings all appeal to \textsc{Similarity}, the analogue behaviour of entropy and horizon area. But Bekenstein also uses \textsc{Preservation} to argue for a black hole entropy.\footnote{It should be noted that there is a tacit assumption of information being conserved.} If a violation of the second law is to be avoided, black holes need to have an entropy. Bekenstein illustrates this intuition by considering the following example. Stellar objects have an entropy. Once they reach the end of their lives, some of them collapse into black holes. If black holes didn't have an entropy, entropy effectively would have been destroyed during the formation of a black hole. This however constitutes a violation of the second law. It is this argument that lead Bekenstein to the fomulation of the generalised second law, which was already introduced in the previous section and which states that ``The common entropy in the black hole exterior plus the black hole entropy never decreases'' \citep[p.2339]{bekenstein_black_1973}.

We will now move on to yet another founding father of black hole thermodynamics, Stephen Hawking, and briefly introduce the argument he made in favour of a black hole entropy.

\subsubsection*{Hawking and thereafter}

Stephen \citet{hawking_black_1976} adopts some of Bekenstein's reasoning, but in addition provides his own argument for showing that the black hole entropy equals the statistical mechanical entropy. As opposed to Bekenstein, however, Hawking himself does not explicitly draw a distinction between the thermodynamic and the statistical entropy. He begins by assuming that there exists a finite amount $\sigma$ of initial, uniformly distributed states that may give rise to a particular black hole. Just like in the ordinary statistical case, the entropy should be equivalent to $S_{BH}=\ln \sigma$. The entropy furthermore is required to be a function of $M$, $Q$ and $J$. Hawking then demands that this entity always increases when matter or radiation falls into the black hole and that it is superadditive for two merging black holes. The only functions that satisfy the above criteria, as Hawking finds, are functions of the horizon area, the simplest of which is the area times a constant, which then turns out to be $c^3/4G\hbar$. If matter falls into the black hole, the change in values of $M$, $Q$ and $J$ leads to an increase in $\sigma$ that's at least as large as the old value of $\sigma$ times the number of configurations of the accretting matter. Hawking has therefore given a statistical mechanical derivation of what he takes to be the generalised second law. 

Just like Bekenstein, Hawking takes $\ln \sigma$ to represent an agent's ignorance over the system's underlying micro-configuration. In his conclusion he writes: ``The conclusions of this paper are that there is an intimate connection between [black] holes and thermodynamics which arises because information is lost down the hole.'' \citep[p.179]{hawking_black_1976}. 

Due to Hawking's lack of explicit distinction between the thermodynamic entropy and the statistical mechanical entropy, it remains a matter of speculation whether he considers the two to be distinct\footnote{In the sense of one being the generalisation of the other.} or not. Speaking in favor of him taking black holes to have a thermodynamic entropy is furthermore the fact that he in a later publication describes black holes as having ``thermodynamic properties'' \citep[p.577]{hawking_thermodynamics_1983}, such as temperature and entropy. This suggests that Hawking is following \textsc{Statistical Mechanics}.

Disregarding whether or not Hawking himself distinguished between statistical mechanical and thermodynamic entropy, this distinction has washed out significantly in the black hole literature in subsequent years. Wald for example introduces the generalised second law as ``the total entropy of matter outside black holes plus 1/4 the surface area of all black holes never decreases with time. This suggests that the laws of black hole mechanics literally \textit{are} the ordinary laws of thermodynamics applied to a system containing a black hole'' \citep[p.55, original emphasis]{wald_black_1992}. For Wald, therefore, there is no doubt that the Bekenstein-Hawking entropy is just the ordinary thermodynamic entropy. 
\begin{quote}\small\singlespacing 
[...] we must interpret $S_{bh}$ as representing the \textit{physical} entropy of a black hole, and that the laws of black hole mechanics must truly represent the ordinary laws of thermodynamics as applied to black holes. \citep[p.18, original emphasis]{wald_thermodynamics_2001}
\end{quote}
He nevertheless also asserts that the question of how they arise from the underlying statistical mechanics is a mystery.

\subsection{What the arguments don't show}

The arguments presented above were a mixture of \textsc{Similarity}, \textsc{Preservation} and \textsc{Statistical Mechanics}. We will discuss now in more detail how they do not succeed in establishing that black hole entropy is thermodynamic entropy. 

\textsc{Similarity}, as pointed out earlier, appeals to the similar behaviour of black hole entropy and thermodynamic (or statistical) entropy. Such arguments can at most establish an analogy, but not an identity between the two entropies. Take as an illustration the case of `similar' behaviour of a pendulum and an LC-circuit (made of a conductor L and a capacitor C): both are versions of a harmonic oscillator, but yet nobody would claim that spatial displacement is equal to electrical current. The majority of the arguments given by Bekenstein fall into this category and so they can at most establish an \textit{analogy} between thermodynamics and black holes\footnote{To remain fair, this is all Bekenstein aims to achieve.}.

What about arguments that appeal to the preservation of the second law of thermodynamics? Must we not consider black holes as having a thermodynamic entropy in order to save the second law, one of the most well-established laws in physics? In order to save the second law, the entropy of the black hole must rise by at least the same amount as entropy of its surroundings was `lost' (we may consider the example of a box of gas that falls into the black hole). However, if we do not require our system to return to its original state, then `apparent violations' are indeed nothing unusual in thermodynamics. For example, it is not hard to come up with examples where heat is transferred from a colder to a warmer body, if system and environment are not required to return to their original state. Such examples would however not constitute a violation of the second law. The second law was first introduced, and is best understood, in terms of reversible \textit{cycles}.\footnote{Some more recent approaches to phenomenological thermodynamics, in particular the axiomatic approaches of \cite{lieb_guide_1998} do not require the notion of cycles in order to derive the second law. We do not consider them here.} And so arguments that include thermodynamic objects falling into the black hole are to a certain extend unsatisfactory, as they do not in an obvious manner allow us to construct a cycle. Despite this criticism of the use of \textsc{Preservation} in the presented arguments, \textsc{Preservation} nevertheless will be the strategy we will chose in order to demonstrate that black holes indeed have a thermodynamic entropy that scales with their horizon area. As opposed to previous approaches, however, we will consider quasi-static, reversible cycles. 

It was shown above that Bekenstein was not interested in establishing $S_{BH}=S_{TD}$ but instead wanted to show that the black hole entropy is a statistical mechanical entropy. Hawking equally was interested in deriving black hole entropy from statistical mechanical considerations. Both authors therefore follow \textsc{Statistical Mechanics}. The problem with \textsc{Statistical Mechanics} however is, that it does not allow us to conclude that the derived statistical mechanical entropy indeed is the phenomenological entropy. The identification of thermodynamic entropy with its statistical mechanical generalisations is highly contested \citep{sklar_reduction?_1999,callender_reducing_1999,goldstein_boltzmanns_2001} in the sense that it is far from clear which, if any, statistical mechanical entropy is a suitable candidate for representing the thermodynamic entropy. 

At this point, one might be tempted to question the whole enterprise of identifying black hole entropy with thermodynamic entropy: why should we care about whether black holes have a \textit{thermodynamic} entropy that is proportional to their surface area? Statistical mechanics is more fundamental than thermodynamics, why not focus instead on the problem of deriving a statistical mechanical underpinning? As important as this last task is, there are nonetheless things to say in favor of establishing black holes having a thermodynamic entropy. 

First, it is after all thermodynamics that gives physical meaning to temperature, heat and work and of course entropy. One of the tasks of statistical mechanics is to recover the laws of thermodynamics, but its range of application is naturally much broader than that. It is possible to assign a statistical mechanical entropy to the whereabouts of my bike keys, but this entropy is void of any (important) physical meaning. It tells me nothing about the thermal properties of my house. In particular, it does not allow me to talk about our familiar understanding of heat and temperature. Second, whereas in the classical case the microstates and the dynamics underlying the statistical mechanics is well known, the situation is much more complicated in the case of black holes. Here, a statistical mechanical description of the black hole is anything but business as usual and ought to be speculative in nature. It is little condolence that string theory has derived the Bekenstein-Hawking formula for black holes \citep{strominger_microscopic_1996}, given that it is unclear what physical meaning we ought to assign to it. In short: given that the statistical mechanical underpinning is so remarkably unclear (see \citep{bekenstein_bekenstein-hawking_2008} for a  summary of all the attempts to derive the black hole entropy), it is a good idea to at least establish that black holes are thermodynamic in the first place.

\section{Black Holes as Thermodynamical Objects}

In the following we present an argument which shows that black hole entropy can be considered to be actual, genuine, well-behaved thermodynamic entropy, given a range of external parameters. We do so by considering a Carnot cycle with a black hole coupled to a photon gas as the working medium. In the next section, we will discuss what motivates such an approach. 

\subsection{Motivation}

\textit{What is new?}

The approach discussed here tries to avoid statistical notions, and in particular the concept of `information', as much as possible. The goal is to show that black holes can indeed be considered to be true thermodynamics objects and to investigate whether their entropy is given by the Bekenstein-Hawking formula. To do so, the black hole will be coupled to an object which is uncontroversially thermodynamic in the sense that it behaves according to the laws of thermodynamics and that it can be described by the usual thermodynamic parameters. A photon gas at temperature $T_g$ will take the role of this system. It will then be shown that the joint system can be used as the working substance of a Carnot heat engine. This approach differs from \textsc{Similarity}, which considers the behaviour of isolated black holes. Instead, we say: if it really is a well-behaved thermodynamic system, then it must interact with other thermodynamic systems like a thermodynamic system does and both of them together must behave like a thermodynamic system\footnote{See \cite{curiel_classical_2014} for a justification of arguments of this sort.}. 

There are a few assumptions necessary for the argument. One of them is that the second law of thermodynamics holds of our combined system. This is far from trivial and follows along the lines of \textsc{Preservation}. We believe that it is nevertheless justified to do so, given that we show the existence of stable equilibrium states which allow us to consider reversible, quasi-static \textit{cyclic processes}, such as first considered by Kelvin and Clausius, as will be discussed in the following.
\\[2em]
\textit{Why cycles?}

It was cyclic processes that inspired (or rather defined) the second law in the first place. Historically, the second law was formulated in terms of the efficiency of heat engines (Carnot): no cyclic process is more efficient than a reversible process. This emerges from the (phenomenological) fact that heat naturally always flows from warm to cold and never the other way round. From this it also follows that one cannot have a heat engine that (operating in a cycle) takes heat from a reservoir and transforms it into work without producing any excess heat. As the efficiency of a reversible heat engine furthermore does not depend on the nature of the working fluid but is a function of the temperatures of the involved heat reservoirs only, one can define an absolute temperature scale.  

Our main point of interest, thermodynamic entropy, enters the picture only now. From the Kelvin statement, one derives the so-called \textit{Clausius inequality} $\oint \dbar Q/T\leq 0$ with equality for reversible cycles, where $\dbar Q$ is the heat flux into the system and $T$ is the thermodynamic temperature of the heat reservoirs (the above equation is the limiting case of the discrete Clausius inequality $\sum_i \dbar Q_i/T_i$, for which it is more obvious that the $T_i$ indeed correspond to the temperatures of the involved heat reservoirs from which heat is extracted). The derivation of the Clausius inequality involves the cyclic process of a motor, which is instantiated by a series of Carnot heat engines that drive the motor, all operating between a principal heat reservoir and a number of auxiliary heat reservoirs. The motor will be back to its original state after a full cycle and together with the Kelvin statement of the second law, the Clausius inequality is derived. The temperatures occurring in the Clausius inequality all refer to the temperatures of the heat reservoirs. As $\int \dbar Q/T$ is path independent, one can now define a state function, the entropy function, whose value (up to an arbitrary constant) solely depends on the state of the system. The Clausius inequality then becomes the entropy version of the second law $\Delta S\geq 0$ with equality for reversible processes in a thermally isolated system. The entropy can only be determined up to a constant. 

In the approach given here, we will show that black holes behave like thermodynamic objects with a thermodynamic entropy given by the Bekenstein-Hawking formula. To do so, we will first show that a black hole coupled to a photon gas and enclosed in a box behaves like a thermodynamic object insofar as that it can be used as a heat engine and perform a Carnot cycle. From the above discussion we can distill a few requirements that such a system necessarily must fulfill. First, each step must take place on a reversible path. Whether or not such reversible transitions are possible hinges crucially on the existence of equilibrium states: for a process to be reversible, the system needs to undergo quasi-static changes, at each instant of which the system is at equilibrium. A necessary requirement for a system to be able to undergo a reversible cycle is hence that it can be at thermal equilibrium for a range of external parameters. Furthermore, the existence of ideal heat baths is needed. It is with the temperature of the heat baths that we derive the absolute temperature scale.

It will shortly be shown that for the combined system, black hole, radiation and box, a thermodynamic temperature exists, but: does it follow that the black hole itself has a temperature? If we take Hawking's famous result for granted, then a black hole emits radiation with a thermal spectrum corresponding to some temperature $T_{BH}$. It is then possible for the black hole and the photon gas to be in a stable equilibrium state for which $T_{BH}=T_g$, as will be demonstrated. 
This potential for being in equilibrium with a thermodynamic object that \textit{has} a temperature and being able to undergo reversible changes, allows us to say that the black hole equally \textit{has} a temperature and not merely \textit{emits} at a certain temperature. However, it should be mentioned that even though the black hole can be at equilibrium with the photon gas, the transitivity relation is restricted in the case of black holes. As \citet{hawking_black_1976} remarked: a naked black hole could not be in stable equilibrium with an infinite heat bath due to its negative heat capacity. Even if the two started out in equilibrium, fluctuations would quickly lead to a run-away process, as will be explained in more detail shortly.

Let us briefly summarise the above: we will show that a black hole enclosed in a box with radiation gas can be used as a working substance for a Carnot cycle. Whereas we may expect a box with a piston and filled with gas to trivially be usable as a heat engine, it is not at all obvious whether we can expect the same when we include a strange stellar object such as a black hole. The main requirement for success is that the black hole can be in stable equilibrium with the gas and undergo a series of quasi-static transitions. It will be shown now, that this is indeed possible for a certain range of external parameters. 

\subsection{Equilibrium Conditions for Black Holes and Photon Gases}

Black holes have a negative heat capacity of order $\propto -M_{BH}^2$, which contributes to a very counter-intuitive behaviour in thermal contexts. This means, that upon absorbing heat (or energy in general), black holes \textit{cool down}. The first problem with having a negative heat capacity is that we cannot expect systems that start out at different temperatures, to ever be at equilibrium with each other. The second problem is that even if two systems start out at equilibrium, this equilibrium may not be robust against small fluctuations. As an illustration, consider a black hole initially in thermal equilibrium with a surrounding, infinite heat bath. Due to a fluctuation, its mass increases slightly, which in turn leads to a decrease in temperature. As its absorption cross section has now increased, it absorbs even more photons from the surrounding infinite heat bath. At the same time, as it has cooled down, its rate of emission decreases, and so the black hole will become even bigger and colder, and so on. The converse situation, in which the black hole fluctuates towards a higher temperature and a smaller mass, works analogously, leading to the complete evaporation of the black hole. It was Hawking who first expressed this worry and arrived at the conclusion that ``black holes cannot be in stable thermal equilibrium in the situations in which there is an indefinitely large amount of energy available.'' \citep[p.2]{hawking_black_1976}. 

Even though black holes cannot be in stable equilibrium with an infinite heat bath, they can be in stable equilibrium with a photon gas and enclosed in a box, for a certain range of parameters\footnote{A similar discussion can be found by \citet{custodio_thermodynamics_2003}.}. This then allows us to have the joint system undergo quasi-static transitions. Due to the negative heat capacity of the black hole, there nevertheless do remain some subtleties, which will discussed shortly. Placing the black hole in a box also allows us to extract work from the system in the old-fashioned way by having a piston which allows the volume of the box to vary or to be varied. 

To begin, we take the box to be completely isolated with a constant total energy of 
\begin{equation}
E_{tot}=E_g+E_{BH}, \label{eq:totalenergy}
\end{equation}

where $E_g=\alpha V T_g^4$ is the energy of the photon gas within the box of volume $V$ and at temperature $T_g$ and $\alpha=(\pi k^2)^2/(15c^3\hbar^3)$. $E_{BH}=Mc^2$ the energy of the black hole with mass $M$.

We re-arrange equation (\ref{eq:totalenergy}) in such a way that the temperature of the photon gas is a function of the total energy, the black hole mass and the volume of the box:
   
  \begin{equation}\label{eq:tempgas}
  T_g= \left[\frac{(E_{tot}-Mc^2)}{\alpha V}\right]^{1/4}
  \end{equation}
  
  Hawking found that black holes radiate at the same rate as an ordinary body would if it were at a temperature inversely proportional to the black hole mass, namely at \citep{hawking_particle_1975}

  \begin{equation}\label{eq:tempbh}
  T_{BH}=\frac{\hbar c^3}{8\pi kG}\frac{1}{M}.
  \end{equation}


 At equilibrium, we require the black hole and the photon gas to be at the same temperature, $T_{eq}=T_{BH}=T_g$. Equating equation (\ref{eq:tempgas}) and (\ref{eq:tempbh}) and rearranging the terms leads to
  
  \begin{equation}
f(M)=M^4(M-E_{tot}/c^2)+\beta V\label{eq:quintic}=0,
  \end{equation}
  
 where for convenience we introduce the constant $\beta=\frac{\hbar c^7}{15(8)^4 \pi^2 G^4}$, (notably, this is not temperature). We can consider the above equation to be the \textit{equation of state} for the system, black hole and photon gas. Equation (\ref{eq:quintic}) is equivalent to demanding that $\frac{\partial S_{tot}}{\partial M}=0$ with $S_{tot}=S_{BH}+S_g$, but at this stage we want to avoid any entropy talk as much as possible.
  
Being a quintic equation, it's not always possible to analytically find the roots of $f(M)$. However, fortunately it is still possible to extract a sufficient amount of information from equation (\ref{eq:quintic}) that allows one to characterise the equilibrium conditions for the joint system. 

Firstly, we can easily see that $f(M)\rightarrow \pm \infty$ as $M\rightarrow \pm \infty$. In addition, $f(M)$ has two turning points, namely at $M=0$ and $M=\frac{4}{5}E_{tot}$, where we set $c=1 \frac{m}{s}$ for simplicity. Having two turning points means that $f(M)$ can have at most three roots. At zero mass, $f(M=0)=\beta V>0$, and since we know that for $M\rightarrow -\infty$, $f(M)\rightarrow -\infty$, one root must be at negative mass and hence irrelevant. Only the interval of $0\leq M\leq E_{tot}$ is physical. 

The existence of the two other equilibrium points depends on whether or not $f(M=\frac{4}{5} E_{tot})$ is positive or negative (we require it to be negative), which can be rewritten as the following inequality\footnote{These values are consistent with those of \citet{hawking_black_1976}, who arrives at the same conclusion but with a statistical approach.}:

\begin{equation}
\frac{V}{E_{tot}^5}\leq \frac{0.082}{\beta} \label{eq:inequality}.
\end{equation}

Equation (\ref{eq:inequality}) provides an upper bound for the volume of the box, given a fixed total energy (or alternatively a lower bound for the overall energy, given a fixed volume). This upper limit to the box size turns out to be sufficiently large: for a black hole of mass $M=M_\odot$ that accounts for $4/5$ of the total energy the box may be as large as $1.5\times 10^{30}m$. 

If the volume $V$ exceeds the limiting volume $V_l$, the only equilibrium that could be achieved is at $M=0$, in which case the box contains only radiation but no black hole. Intuitively this makes sense: if we place a black hole into a box which is too large compared to the total energy, black hole and gas could therefore never equilibrate. If at some initial time $T_{BH}>T_g$, the black hole would absorb energy, grow and cool down, but the gas would also cool down and at a faster rate such that their temperature would never be equal. For $T_{BH}<T_g$ the opposite case would hold, leading to the complete evaporation of the black hole. 

\begin{figure}
\centering
\fbox{\includegraphics[width=.7\textwidth]{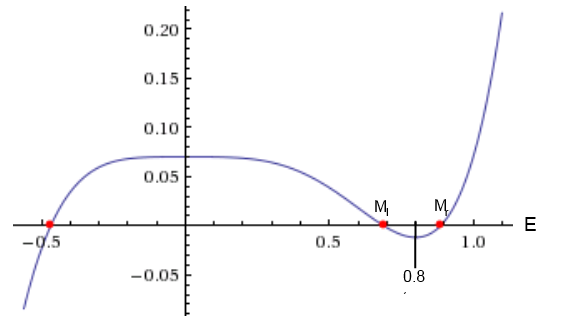}}
\caption{Exemplary graph for $f(M)$. The units on the $x$-axis are set to units of $E_{tot}$. The box volume used for this graph is $V=0.07$, as can be seen from $f(M=0)=0.07$. We can see that two roots exist left ($M_l$) and right ($M_l$) of the turning point $M=0.8 E_{tot}$. For a volume larger than allowed by equation (\ref{eq:inequality}), these two roots don't exist. The root left on the far left is taken to be unphysical, as it refers to a negative black hole mass.}
\label{fig:EoS}
\end{figure}

In Figure (\ref{fig:EoS}) we see two equilibrium points, one left ($M_l$) and right ($M_r$) of $M=\frac{4}{5}E_{tot}$. It is now essential to see whether those are stable and to show that a temperature difference between gas and black hole, $T_g<T_{BH}$ or $T_g>T_{BH}$ does not lead to a runaway process of the kind described above. One way to do so is by considering $f(M)=dS/dM$ and thereby interpreting the roots $M_l$ and $M_r$ as local extrema of the entropy function and fixing the energy and volume \citep{custodio_thermodynamics_2003,page_particle_1976} . It can then be shown that since $f'(M<\frac{4}{5}E)<0$ and $f'(M>\frac{4}{5}E)>0$, the root at $M_l$ must be a stable local entropy maximum. The stability criterion can also be directly read off the graph in Figure (\ref{fig:EoS}).

This approach, however, presupposes the existence of an entropy function, which is what we want to avoid. It is still possible to show that the left equilibrium point, $M_l$, is stable by considering the rate of change in temperature upon a change in black hole mass. If it is the case that 
\begin{equation}
\frac{dT_{BH}}{dM}<\frac{dT_{g}}{dM},
\end{equation}

then the gas can react quickly enough to any fluctuations in the black hole and there will be no run-away processes. Inserting all the relevant entities into the above equation, we get

\begin{align}
-\frac{\hbar c^3}{8\pi Gk}\frac{1}{M^2}&<-\frac{1}{4\alpha V}\left[\frac{\alpha V}{E_{tot}/c^2-M}\right]^{3/4},\\
\frac{\left(E_{tot}/c^2-M\right)^3}{M^8}&>\frac{1}{4^4\beta V}.
\end{align}

Applying the previous derived constraint for the box volume in the above inequality, it turns out that the condition is fulfilled for $M<4/5E_{tot}$ and so it is the left root $M_l$ of $f(M)$ that denotes a stable equilibrium state.


We have therefore shown that (given a restriction on the ratio between total energy and box volume), a black hole can be in stable equilibrium with a surrounding photon gas\footnote{Many thanks to David Wallace for suggesting a shorter derivation based on Thirring's stability condition \citep{thirring_systems_1970}. For two systems $A$ and $B$ to be at thermal equilibrium, the following stability condition must be fulfilled: $\frac{c_A c_B}{c_A+c_B}>0$, where $c_i$ is the heat capacity of the respective system. If $c_A<0$ and $c_B>0$ then a necessary requirement for the two systems to be at thermal equilibrium is that $c_B<\vert c_A \vert$. This is another way to derive the upper limit to the box size.}, within the above defined parameter range. This means that for small perturbations, black hole and photon gas quickly equilibrate themselves again. For a Carnot cycle, transitions are taken to take place quasi-statically, and so at each time step the system has time to re-equilibrate itself.

With the above reassurance that a black hole can be in equilibrium with a photon gas when enclosed in a box of certain maximum volume, we now finally proceed with modeling a Carnot cycle. We will furthermore recover the well known expression for the black hole entropy from this analysis.  
  
  \subsection{Modelling a Black Hole Carnot Cycle}
  
\subsubsection{Setup}

Just as above, we take as a working medium a black hole surrounded by a photon gas, enclosed in a box and attached to a piston. Both are at thermal equilibrium with each other at all times, namely $T=T_{BH}=T_g$. 

 The total energy of the system is given by 
 \begin{equation}
 E_{tot}=U=Mc^2+\alpha VT^4, \label{eq:energy}
 \end{equation}
 where $M$ is the mass of the black hole, $V$ the volume of the box and $\alpha=\frac{\pi^2k^4}{15c^3\hbar^3}$ the usual radiation constant. One of the assumptions made earlier was that a photon gas can be considered a true thermodynamic object which is described by thermodynamic variables. It therefore has an entropy, which is
 
 \begin{equation}
 S_{rad}=\frac{4}{3}\alpha V T^3,
 \end{equation}
 and exerts a pressure on the walls of the box of
 \begin{equation}
 P_{rad}=\frac{1}{3}\alpha T^4.
 \end{equation}
 
 The box, containing both radiation and black hole, can be brought in contact with one of two heat baths at temperatures $T_1$ and $T_2$, where $T_1<T_2$. 
  
\begin{figure}
\centering
\fbox{\includegraphics[width=.7\textwidth]{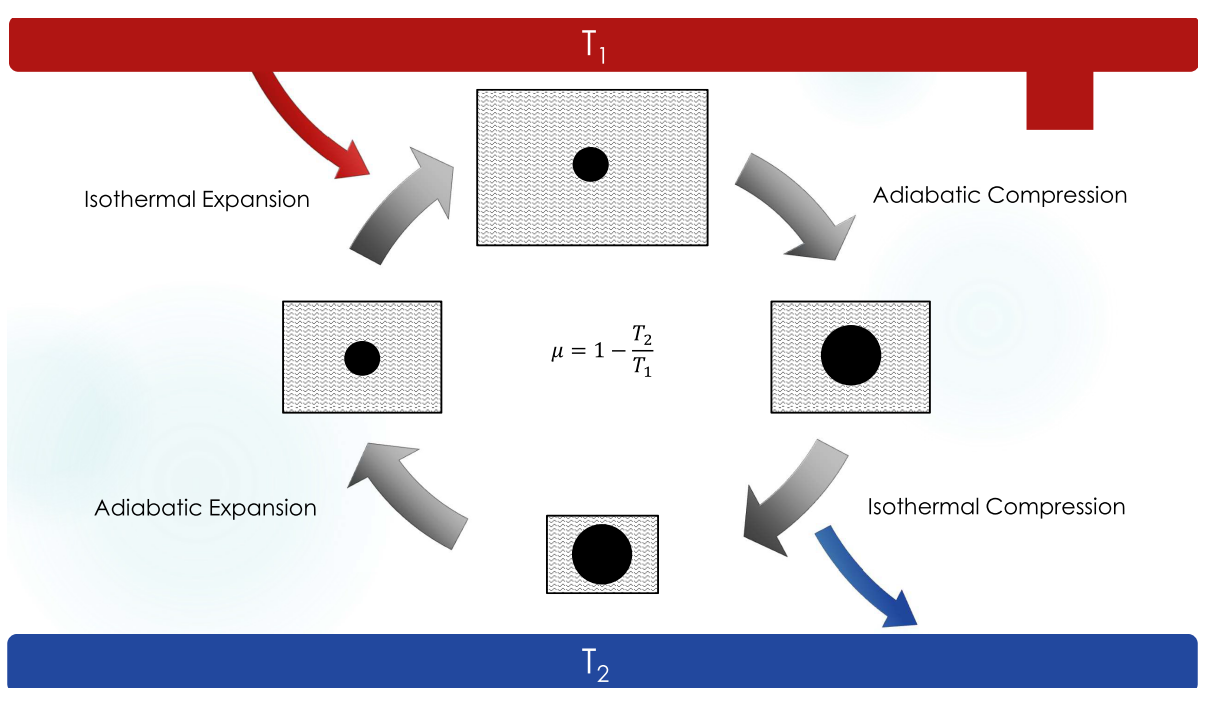}}
\caption{Schematic illustration of a black hole Carnot cycle. The system consists of a black hole and a photon gas, enclosed in a box. The size of the black hole is proportional to the temperature of the system, i.e. small is hot and large is cold. }
\label{fig:carnot}
\end{figure}

We now let the system go through the standard\footnote{Strictly speaking it is a reverse Carnot cycle. The standard Carnot cycle would be isothermal expansion, adiabatic expansion, isothermal compression, adiabatic compression. Since the system has negative heat capacity, however, this standard Carnot cycle would serve as a refrigerator. The reverse Carnot cycle on the other hand acts as a heat engine, which is why we consider the reversed version.} Carnot cycle, which consists of isothermal expansion, adiabatic compression, isothermal compression and adiabatic expansion. An illustration of the process is given in Figure \ref{fig:carnot}.

  \subsubsection{Isothermal Expansion}

Since the overall system has a negative heat capacity, the notion of isothermal expansion during the Carnot cycle becomes a more subtle business than in the case of a system with positive heat capacity. In the latter case, a system is brought in contact with the hotter heat reservoir, and kept at the same temperature while expanding. Here, given its negative heat capacity, the system will be brought to a temperature just \textit{below} the hot reservoir and kept there. When a small amount of heat flows from the reservoir to the system, the system will cool down very slightly. This is counterbalanced by simultaneously expanding the box, for which the system will need to do work, thereby heating up (due to its negative heat capacity). This way it is possible to expand the system from $V_1$ to $V_2$ and keep it at the same temperature, slightly below that of the heat bath. Since this temperature difference can be made arbitrary small, we still have effectively $T_{BH}\approx T_{1}$. 

During this isothermal expansion the volume of the box changes from $V_1$ to $V_2$. The first law of thermodynamics
  
  \begin{equation}
  \Delta U_{12}=Q_{12}+W_{12}
  \end{equation}
  
holds ($W_{12}$ is the work done on the system) and so we can calculate the heat flowing from the hot reservoir as
\begin{equation}
\dbar Q_{12}=dU+PdV=\left(\frac{\partial U}{\partial T}\right)_VdT+\left(\frac{\partial U}{\partial V}\right)_T dV + PdV
\end{equation}

The black hole mass only depends on the temperature of the black hole, and not on the volume of the box. Therefore, with the hot reservoir keeping the system at constant temperature $T=T_1$, we obtain for the total heat flux during the isothermal expansion of the box:
\begin{equation}
Q_{12}=\int_{V_1}^{V_2} \left(\alpha T_1^4 +\frac{1}{3}\alpha T_1^4\right)dV=\frac{4}{3}\alpha T_1^4\left(V_2-V_1\right).
\end{equation}

The total system's entropy, which we have allowed to exist, since we must assume that the second law applies to the box system as a whole, must change during this process by 
\begin{equation}
\Delta S_{12}=\int_1^2 \frac{\dbar Q}{T}=\frac{4}{3}\alpha T_1^3(V_2-V_1).
\end{equation}

Only the photon gas performs work and since its pressure only depends on the temperature, the process is not only isothermal but also isobaric and we obtain for the work term
\begin{equation}
W_{12}=-\int_{V_1}^{V_2} P_{rad}dV=-\frac{1}{3}\alpha T_1^4 (V_2-V_1).
\end{equation}

The total change of energy of the system during this process is then




\begin{equation}
\Delta U_{12}=c^2 \Delta M_{12}+4\alpha \left(V_2T_1^4-V_1T_1^4\right)=\alpha T_1^4(V_2-V_1),
\end{equation}

  \subsubsection{Adiabatic Compression}
  
During the adiabatic compression, the system does not exchange energy in the form of heat with its surroundings. The volume changes from $V_2\rightarrow V_3$, but no heat is exchanged with the environment, $Q_{23}=0$. The compression requires work, and so the total energy of the system changes and the system's temperature drops from $T_1$ to $T_2$. As the black hole mass is temperature dependent, its mass will now increase, $M_1\rightarrow M_2$. 

The adiabatic condition for a process with no heat exchange is given by:
\begin{equation}
\Delta U_{23}=W_{23}.\label{eq:adiabatic}
\end{equation}

Using equations (\ref{eq:energy}) and (\ref{eq:adiabatic}), we obtain the work done on the system during the adiabatic compression:
\begin{equation}
W_{23}=\gamma c^2\left(\frac{1}{T_2}-\frac{1}{T_1}\right)+\alpha \left(V_3T_2^4-V_2T_1^4\right),\label{eq:workadiabatic}
\end{equation}

where we used the Hawking expression relating the temperature and mass of a black hole by $M(T)=\frac{\hbar c^3}{8\pi G kT_{BH}}=\frac{\gamma}{T_{BH}}$.

The first term of the above equation is $\geq 0$ and the second term $\leq 0$. 

The work itself is only performed against the pressure of the photon gas, but the change of internal energy affects both the gas and the black hole at the center of the box. These must be equal, and so we differentiate equation (\ref{eq:energy}) and equate it with the work done on the photon gas:

\begin{equation}
dU=c^2 dM +4\alpha VT^3 dT+\alpha T^4 dV=-\frac{1}{3}\alpha T^4 dV.
\end{equation}

Re-arranging the terms then leads to the following condition:
\begin{equation}
c^2 dM+4\alpha VT^3 dT+\frac{4}{3}\alpha T^4 dV=0 \label{eq:differentiated}
\end{equation}

The mass of the black hole only depends on the temperature, $M=M(T)$ and so $dM=\left(\frac{\partial M}{\partial T}\right)_VdT$. Equation (\ref{eq:differentiated}) after some re-arranging of the terms then yields:

\begin{align}
\frac{dV}{dT}+\frac{3V}{T}=-\frac{3c^2}{4\alpha}\frac{\partial M}{\partial T}\frac{1}{T^4}.\label{eq:prediff}
\end{align}

If we now insert the concrete expression for the black hole mass, $M(T)=\gamma/T$, equation (\ref{eq:prediff}) becomes
\begin{equation}
\frac{dV}{dT}+\frac{3V}{T}=\frac{3\gamma c^2}{4\alpha}\frac{1}{T^6}.
\end{equation}

We solve this equation for the temperature-dependent volume, and so the above equation becomes

\begin{equation}
V(T)=-\frac{3\gamma c^2}{8\alpha T^5}+\frac{K}{T^3},
\end{equation}

where $K$ is a constant. We can re-arrange the above in a useful way which resembles a bit more the traditional way of expressing adiabats:

\begin{equation}
VT^3+\frac{3\gamma c^2}{8\alpha T^2}=\text{constant}.\label{eq:adiabat}
\end{equation}

Equation (\ref{eq:adiabat}) describes the adiabats of the system. These are the paths of no heat exchange with the environment the state of the system follows during an adiabatic transition. Notably, they differ from the adiabats for a pure photon gas by the second term in equation (\ref{eq:adiabat}).

Given that there is no heat exchange with the environment, the entropy change of the total system must be zero.

   \subsubsection{Isothermal Compression}
   
  The isothermal compression stage takes place analogously to the isothermal expansion stage, but this time the system is held at a temperature slightly \textit{above} the temperature of the cold heat bath. 
  
  The amount of heat released is given by
  
  \begin{equation}
  Q_{34}=\frac{4}{3}\alpha T_2^4(V_4-V_3),
  \end{equation}
  
  with an associated entropy change of the total system of $\Delta S_{34}=\frac{4}{3}\alpha T_2^3(V_4-V_3)$. The work performed on the system is

  \begin{equation}
  W_{34}=-\frac{1}{3}\alpha T_2^4(V_4-V_3),
  \end{equation}
  
  and the total change in internal energy is
  
  \begin{equation}
  \Delta U=\alpha T_2^4(V_4-V_3).
  \end{equation}
  
   
  \subsubsection{Adiabatic Expansion}\label{sec:adiabatic}
  
  For the adiabatic expansion $V_4\rightarrow V_1$, the adiabatic relations are given by
  \begin{equation}
V_4T_2^3+\frac{3\gamma c^2}{8\alpha T_2^2}=V_1T_1^3+\frac{3\gamma c^2}{8\alpha T_1^2}.  \end{equation}
  
  Together with equation (\ref{eq:adiabat}) we obtain the following relation:
  
  \begin{equation}
  V_3T_2^3-V_2T_1^3=V_4T_2^3-V_1T_1^3.
  \end{equation}
  
  The work done by the system during the adiabatic expansion is given by
  
  \begin{equation}
       -W_{41}=-\frac{\gamma c^2}{4}\left(T_1^2-T_2^2\right) -\alpha K\left( T_1-T_2\right).
  \end{equation}
  
  \subsubsection{Efficiency}
  
  The efficiency of a Carnot cycle is given by
  
  \begin{equation}
  \mu=\frac{W_{tot}}{Q_{12}}=\frac{W_{12}+W_{23}+W_{34}+W_{41}}{Q_{12}}.
  \end{equation}
  
 The work terms of the adiabatic expansion and compression cancel each other out. We now derive the well-known Carnot efficiency relations by making use of the previously derived adiabatic relations $T_2^3V_3-T_1^3V_2=T_2^3V_4-T_1^3V_1$ in the third step:
 
 \begin{align}
 W_{tot}=Q_{12}+Q_{34}	&=\frac{4}{3}\alpha \left[T_1^4(V_2-V_1)+T_2^4(V_4-V3)\right]\\
 				        &=\frac{4}{3}\alpha \left[T_1\left(T_1^3V_2-T_1^3V_1\right)+T_2\left(T_2^3V_4-T_2^3V_3\right)\right]\\
                        &=\frac{4}{3}\alpha \left[T_1\left(T_1^3V_2-T_1^3V_1\right)+T_2\left(T_1^3V_1-T_1^3V_2\right)\right] \label{eq:makeuse}\\
                        &=\frac{4}{3}\alpha \left[\left(T_1-T_2\right)\left(T_1^3V_2-T_1^3V_1\right)\right]\\
                    	 &=\frac{4}{3}\alpha \left[T_1^4\left(V_2-V_1\right)\left(1-\frac{T_2}{T_1}\right)\right].
                        \end{align}
  
Together with $Q_{12}=\frac{4}{3}\alpha T_1^4(V_2-V_1)$ we obtain an efficiency of

\begin{equation}
\mu=1-\frac{T_2}{T_1}.
\end{equation}

This is the desired efficiency that we would expect from a Carnot engine. Our system thereby really can be used as a working substance for a heat engine.

\subsection{Black Holes and Entropy} \label{sec:BHentropy}

So far, it has been shown that black holes can be in equilibrium with a photon gas when enclosed in a box of certain maximal volume and furthermore, that the whole system can be used as working substance for a Carnot cycle by undergoing a series of quasistatic changes. In this section we will show how it is possible to calculate the black hole entropy just from these considerations (and without having had to make any assumptions about the thermal nature of a black hole beforehand). 
In the present case, a black hole was coupled to another thermodynamic system, a photon gas.
To see why this is relevant, we consider again the entropy changes of the combined system. $S_{tot}$ changed during the isothermal processes $1\rightarrow 2$ and $3\rightarrow 4$ but remained constant during the adiabatic processes $2\rightarrow 3$ and $4\rightarrow 1$. During the isothermal processes, the black hole did not change its state, and so the entropy change must have occurred exclusively in the photon gas.

During the adiabatic processes, however, $S_{tot}$ remained constant, as no energy had been exchanged with the system environment. Nevertheless, between the black hole and the photon gas, energy \textit{must have} been exchanged during the adiabatic transformations. This can be seen by comparing the adiabats of a photon gas in a box with those of our system comprising both photon gas and black hole: in the former case, the adiabats are given by $VT^3=constant$. In the presence of a black hole, however, they are given by $VT^3+3\gamma c^2/4\alpha T^2$. This means that the thermodynamic path of no heat exchange of a solitary photon gas differs from that of our system. As a consequence, during the adiabatic compression, there \textit{must} have been a heat flux out of the photon gas. And since the combined system is isolated and energy is conserved, the heat flux must be into the black hole. This sounds very much like  Bekenstein's argument, but with one crucial difference: we have shown that black hole and photon gas can undergo a \textit{reversible, quasi-static} cycle. This legitimises us in making use of the entropy formula $\Delta S=\int \dbar Q/T$. 

The entropy of the photon gas during such adiabatic compression therefore must \textit{decrease}. Consequently the black hole needs to experience an \textit{increase} of entropy. If the entropy is additive, and we believe that due to the reversible nature of the process this is a valid assumption, they cancel each other out exactly:


\begin{equation}
\Delta S_{Sys,23}=\Delta S_{rad,23}+\Delta S_{BH,23}=0.
\end{equation}  
  


With the help of the adiabatic relations in equation (\ref{eq:adiabat}), we can calculate the entropy change in the photon gas to be
\begin{align}
\Delta S_{rad,23}	&=\frac{4}{3}\alpha \left( V_3T_2^3-V_2T_1^3\right)\\
					&=\frac{4}{3}\alpha\left( V_2T_1^3+\frac{3 \gamma c^2}{8\alpha T_1^2}-\frac{3\gamma c^2}{8	\alpha T_2^2}-V_2T_1^3\right)\\
                    &=\frac{\gamma c^2}{2}\left(\frac{1}{T_1^2}-\frac{1}{T_2^2}\right), \label{eq:gasentropy}
\end{align}

with the usual $\gamma=\frac{\hbar c^3}{8\pi G k}$.

For the above equation, we assumed that $\left(\frac{\partial M}{\partial T}\right)_V=-\frac{\gamma}{T^2}$ on the basis of Hawking's result for the black hole temperature. For an arbitrary $M(T)$, the change in photon gas entropy would instead read:

\begin{equation}
\Delta S_{rad,gen}=- \frac{\gamma c^2}{2} \int_{T_1}^{T_2} \frac{\left(\frac{\partial M}{\partial T}\right)_V}{T} dT.\label{eq:general}
\end{equation}

If our black hole were not a black hole but instead a perfectly reflecting blob of mass $M(T)=M_{blob}$, the right hand side of equation (\ref{eq:general}) would vanish, resembling a zero entropy change of the photon gas. This is exactly what we would expect for the adiabatic transformation of a non-interacting photon gas.

Returning to our system in question, the entropy change of the photon gas as described in equation (\ref{eq:gasentropy}), must exactly counterbalance the entropy change in the black hole:
\begin{equation}
\Delta S_{BH,23}=\frac{\gamma c^2}{2}\left(\frac{1}{T_2^2}-\frac{1}{T_1^2}\right).
\end{equation}

We can see that, up to a constant, the black hole entropy therefore must be of the form
\begin{equation}
S_{BH}=\frac{\gamma c^2}{2} \frac{1}{T^2}=\frac{\hbar c^5}{16\pi G k}\frac{1}{T^2}, \label{eq:BHentropy}
\end{equation}

which is exactly the well-known expression for the black hole entropy. The horizon area scales with $\frac{1}{T^2}$ and so we have derived a black hole entropy that is proportional to the horizon area.

Given that we have coupled a black hole with an indisputably thermodynamic object and showed that the combined system behaves like a thermodynamic object and given that we have shown that there exist a range of accessible and stable equilibrium states between the subsystems, it seems not to be too much of a leap of faith to say that black holes really can be considered to be thermodynamic objects. For our analysis, we furthermore at no point needed to make reference to `information' or statistical mechanical notions of entropy.

\section{Conclusion}  

By constructing a Black Hole Carnot Cycle, we have shown that for a certain range of parameters, black holes can be taken to have a thermodynamic entropy that coincides with the Bekenstein-Hawking entropy. In particular, no reference to information-theoretic (or, in fact, any statistical mechanical) entropy was required to derive this result. 

Some simplifying assumptions have been made for the derivation of this result, and we ignored a number of complicating factors that may play a role for a physically truthful analysis, such as the effect of gravity both on the photon gas and on the sides of the box. However, we may expect the size of the box strongly to mitigate the significance of those effects. In addition, we only considered Schwarzschild black holes, which are crucial for establishing an equilibrium condition but whose existence in the universe is doubted by many. For other black hole types, for example Kerr black holes, it is not immediately obvious how they could ever be in equilibrium with a photon gas. Still, the analysis presented here may be considered as a first step to a more rigorous treatment of black hole entropy. We have restricted ourselves to phenomenological thermodynamics without making use of statistical mechanical tools and thereby shown that, regardless of the relationship between statistical mechanics and thermodynamics, black holes have a thermodynamic entropy. Since black holes sit at the interface of general relativity and quantum theory, taking a step towards a better understanding of black holes is also taking a step towards a better understanding of what is happening at this interface.


  \newpage

  \newpage
  \bibliography{bibliography}

   \end{document}